\documentclass[aps,pre,showpacs,floats,twocolumn,floats,superscriptaddress,floatfix]{revtex4}

\usepackage{graphicx}
\usepackage{dcolumn}
\usepackage{bm,amsmath}

\newcommand{\e}{{\rm e}}

\newcommand{\XR}{{\langle X(t)^2_R \rangle}}

\newcommand{\XRb}{{X(t)^2_R}}  
\newcommand{\dXRb}{{\delta X(t)^2_{r, R}}}

\newcommand{\XRbwt}{{X^2_R}}  
\newcommand{\dXRbwt}{{\delta X^2_{r, R}}}

\newcommand{\LN}{{\rm LN}}

\begin{document}

\title{Large-scale lognormality in turbulence modeled by Ornstein-Uhlenbeck process}
\author{Takeshi Matsumoto}
\email{takeshi@kyoryu.scphys.kyoto-u.ac.jp}
\affiliation{%
Division of Physics and Astronomy, Graduate School of Science, Kyoto University,\\
Kitashirakawa Oiwakecho Sakyoku Kyoto, 606-8502, Japan.
}%
\author{Masanori Takaoka}
\affiliation{%
Department of Mechanical Engineering, Doshisha University,\\
Kyotanabe, Kyoto, 610-0321, Japan.
}%

\date{\today}

\begin{abstract} 
Lognormality was found experimentally for coarse-grained squared
turbulence velocity and velocity increment when the coarsening scale
is comparable to the correlation scale of the velocity
(Mouri \textit{et al.} Phys.~Fluids \textbf{21}, 065107, 2009).
We investigate this large-scale lognormality by using a simple
stochastic process with correlation, the Ornstein-Uhlenbeck (OU)
process. It is shown that the OU process has a similar large-scale
lognormality, which is studied numerically and analytically. 
\end{abstract}

\pacs{Valid PACS appear here}
\maketitle

\section{\label{s:intro}Introduction}
The lognormal distribution appears in a wide range of natural and 
social phenomena (see, e.g., \cite{lsa}).
In fluid turbulence, it is well-known that the distribution 
is an important consequence of the Kolmogorov's 1962 theory 
for modeling fluctuations of the energy cascade rate across 
scales \cite{k62}.
In this 1962 theory, we have a very clear picture 
why the lognormal distribution was the first candidate of the
cascade fluctuations in his refined phenomenology.
Namely, the energy cascade at high Reynolds number can be
modeled as a multiplicative process consisting of a large 
number of independently and identically distributed random
variables. This large number is important for the central 
limit theorem to be applicable to the logarithm of the
multiplicand.

There is a different example of lognormally distributed variables
in turbulence. Laboratory experiments of turbulent boundary layers 
in 1980's suggest that spanwise separations between the low-speed 
streaks follow the lognormal distribution \cite{nn, sm}.
In this case, the underlying mechanism of the lognormally
distributed streaks is not as clear as in the Kolmogorov 1962
phenomenology
because a multiplicative structure for the streaks is not
found immediately.

In this paper we consider yet another lognormal example in 
turbulence, which was recently found experimentally \cite{mht7, mht}.
In Ref.\cite{mht}, it is
observed that the coarse-grained squared velocity and
squared velocity increments over distance $r$
\begin{eqnarray}
  u^2_R(x_c) &=&
  \frac{1}{R}
  \int_{x_c - R / 2}^{x_c + R / 2}
  u^2(x) d x, \label{u2}\\ 
 \delta u^2_{r,\, R}(x_c) &=&
  \frac{1}{R - r}
  \int_{x_c - R / 2}^{x_c + R / 2}
  [u(x + r) - u(x)]^2 d x \label{du2}
\end{eqnarray}
are lognormally distributed when the averaging scale $R$ is
set to $R \sim L_u$.   
Here $L_u$ is the correlation length defined as the integral scale 
of the velocity correlation function $\langle u(x + r) u(x) \rangle$
and the velocity $u$ can be either longitudinal or transverse velocity 
component.
Since the lognormal property holds when $R$ is comparable to the large scale $L_u$,
this property is called ``large-scale lognormality''.
For the asymptotic regime $R \gg L_u$, as a usual result of the central limit theorem, 
the distribution of the coarse-grained quantities become closer and closer 
to normal (Gaussian) distribution.
Notice that the large-scale lognormality, which we consider here, concerns
the range of the coarse-graining scale $R \sim L_u$, which is different 
from the final Gaussian state in $R \gg L_u$.

The large-scale lognormality was observed in the experiments of 
grid turbulence, turbulent boundary layers, and turbulent jets, 
suggesting its universality \cite{mht}. But again as in the case of 
the streaks in the boundary layer turbulence, 
why the coarse-grained turbulence data are lognormally
distributed is not clear.
Our final goal is to understand its mechanism why
they are so.  

As a first step to this goal, we address the following question:
Is the large-scale lognormality a general property of correlated
random variables, not restricted to large-scale fluctuations of turbulence?
The previous results \cite{mht} lead us 
to recognize that 
this balance between $R$ and $L$ is most important for the lognormality.
In order to answer the question, we use the Ornstein-Uhlenbeck process (OU process) 
as a simplest way to generate correlated random variables with the correlation length $L$ 
and check whether or not the coarse-grained quantities like Eqs.(\ref{u2}) 
and (\ref{du2}) follow lognormal distributions in the range $R / L \sim 1$.
Our numerical results suggest
that the answer to the question is yes.
Then we study, by analytical calculations of moments, further details
on how the OU data become close to the lognormal variables.

We believe that this simple approach using the stochastic process 
has some value since reproducing the large-scale lognormality by 
direct numerical simulations of the Navier-Stokes equations can be
computationally quite expensive. It requires a very large domain 
size as compared with the correlation length.

In statistics, taking logarithm of a positive random variable is
known as a common way of symmetrizing transformation, which makes
the skewness of the transformed variable closer to zero \cite{DasGupta}.
In this language, the large-scale lognormality suggests that,
even with correlation, the log transformation can produce
near Gaussian behaviour successfully already
when the averaging scale $R$ is of the order of the correlation scale $L$.
The present study can be interpreted as a model study how
the log transformation begins to work for the correlated random variable 
by changing the coarse-graining scale.

This paper's organization is the following.  
In Sec.~\ref{s:num}, we present numerical data of the 
OU process reproducing the large-scale lognormality.
We then in Sec.~\ref{s:analy} provide analytical calculations 
of moments of the OU process to study how they are close to
those of the lognormal distribution.
We provide summary and discussion in Sec. \ref{s:dcr}.

\section{Numerical experiment of the Ornstein-Uhlenbeck process}
\label{s:num}
We begin here by listing basic properties of the OU process
(see, e.g., \cite{oubasic} and also \cite{pope} in the context of turbulence modeling).
The process is described by the Langevin equation
\begin{eqnarray}
 \frac{d X(t)}{d t} = - \frac{1}{T_L} X(t) + \sqrt{2\kappa} ~\xi(t),
\label{ou}  
\end{eqnarray}
where the Langevin noise $\xi(t)$ is a Gaussian white noise having
the ensemble-averaged mean and variance
\begin{eqnarray}
 \langle \xi(t) \rangle = 0,\quad
 \langle \xi(t)\, \xi(t') \rangle = \delta(t - t').
  \label{xi}
\end{eqnarray}
The linear addition of the Gaussian uncorrelated noise $\xi(t)$ makes
the solution $X(t)$ to Eq.(\ref{ou}) to be a Gaussian random variable.
However the first term of the right hand side brings temporal correlation.
Namely, by writing the initial position as $x_0$, $X(t)$
is characterized as a correlated-in-time Gaussian random variable 
with the mean and variance
\begin{eqnarray}
 \langle X(t) \rangle &=& x_0 \e^{- t / T_L},\\
 \langle [X(t) - \langle X(t) \rangle ]^2 \rangle &=& \kappa T_L (1 - \e^{-2t/T_L}).
\end{eqnarray}
The correlation function can be calculated analytically as
\begin{eqnarray}
 \big\langle [X(t) - \langle X(t) \rangle][ X(t + s)  - \langle X(t + s) \rangle ] \big\rangle 
  &=& \nonumber \\
 \langle [X(t) - \langle X(t) \rangle ]^2 \rangle  \e^{-|s| / T_L}.
\end{eqnarray}
Therefore the integral scale of the OU process is given as
\begin{eqnarray}
\int_0^{\infty}
 \frac{[\langle X(t) - \langle X(t) \rangle ] [ X(t + s) - \langle X(t + s) \rangle ] \rangle}
 {\langle [X(t) - \langle X(t) \rangle] ^2 \rangle} d s
 = T_L.
\end{eqnarray}

If the large-scale lognormality observed in 
turbulence \cite{mht} is a property of correlated random 
variables, it is then likely that 
the following coarse-grained data of the OU process
\begin{eqnarray}
 X^2_R(t) &=&
  \frac{1}{R} \int_{t - R / 2}^{t + R / 2}
  X(t)^2 d t, \label{x2}\\
 \delta X^2_{r, R}(t) &=&
  \frac{1}{R - r} \int_{t - R / 2}^{t + R / 2}
  [X(t + r) - X(t)]^2 d t, \label{dx2} 
\end{eqnarray}
are lognormally distributed when the averaging scale $R$ 
is in the similar range, such as $ 1 \le R / T_L \le 100$.
Equations (\ref{x2}) and (\ref{dx2}) correspond to the
turbulence quantities Eqs.(\ref{u2}) and (\ref{du2}), 
respectively.

\begin{figure}
\centerline{ 
\includegraphics[scale=0.62]{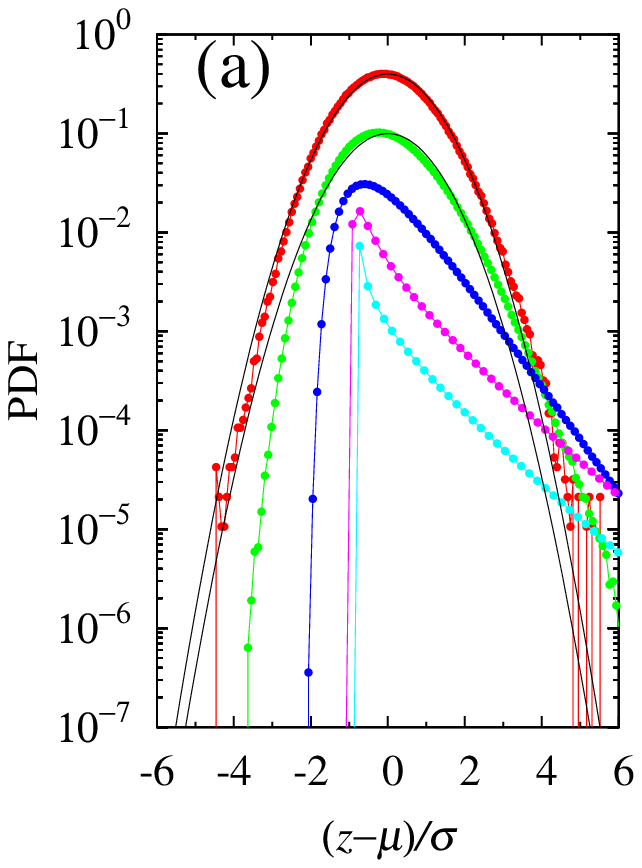}
\hspace{0.2cm} 
\includegraphics[scale=0.62]{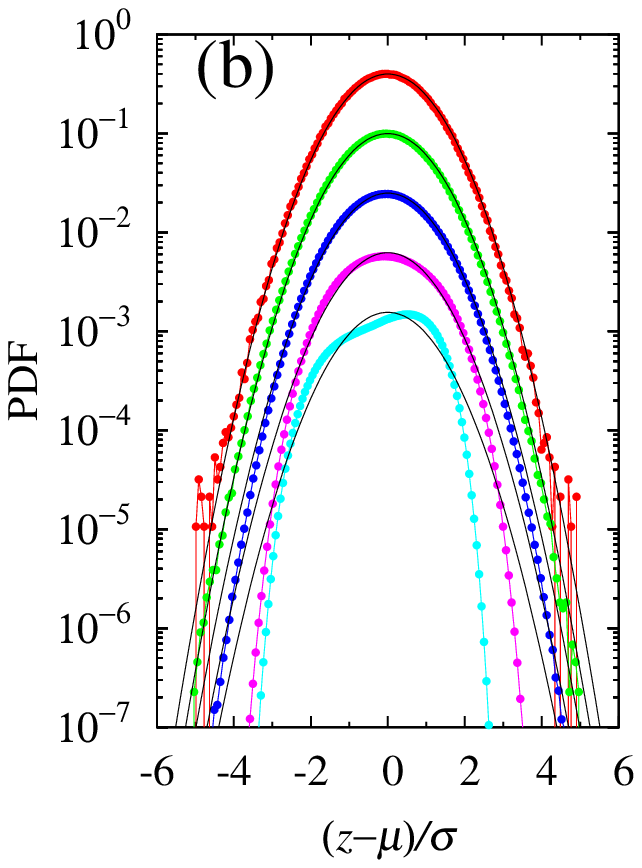} 
} 
\caption{\label{pdf} (Color online) Probability density functions (PDF) of
(a) $X^2_R$ (Eq.(\ref{x2})) and (b) its logarithm
$\ln  X^2_R $. The random variables are normalized to have
zero mean and unit standard deviation.
The averaging scales here are $R = 1000 T_L, 100 T_L, 10 T_L, T_L, T_L/10$, which
correspond to the curves from top to bottom.
The PDFs are shifted vertically by being multiplied by factor $0.25$ 
for clarity. The solid curve denotes the Gaussian distribution.}
\end{figure}
\begin{figure}
\centerline{ 
\includegraphics[scale=0.62]{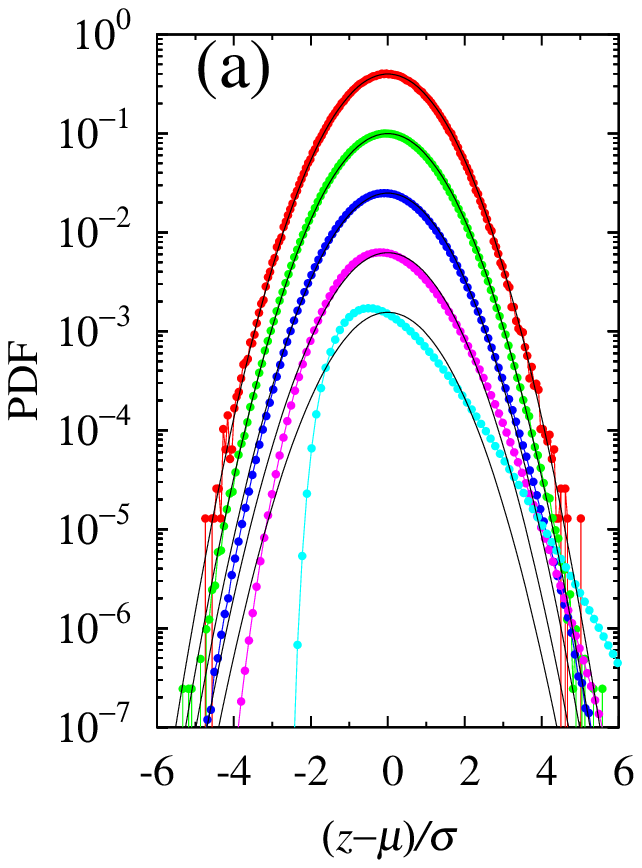}
\hspace{0.2cm} 
\includegraphics[scale=0.62]{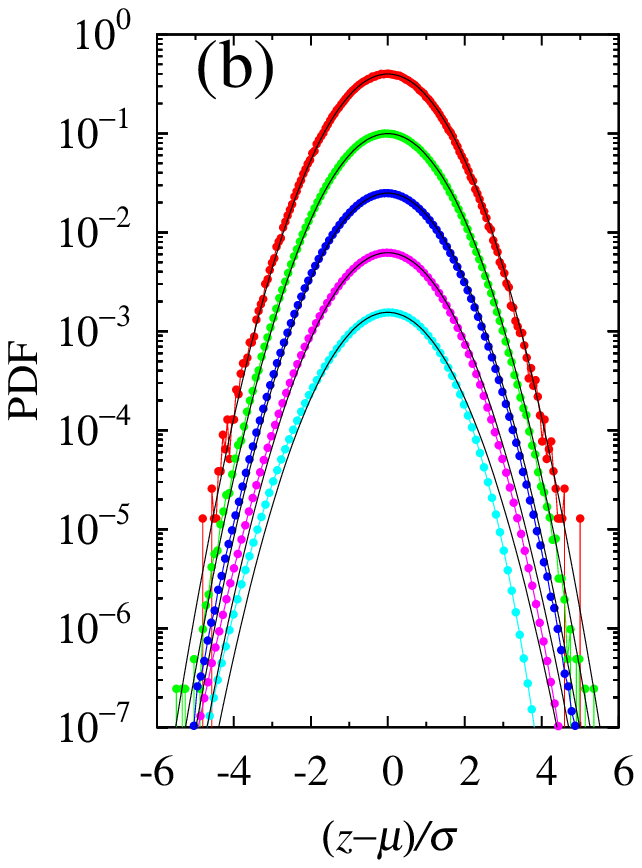}  
} 
\caption{\label{pdfd} (Color online) Same as Fig.~\ref{pdf} but for
 the squared increment
(a) $\delta X^2_{r, R}$ (Eq.(\ref{dx2})) and (b) its logarithm
$\ln  X^2_{r, R}$. Here $r = T_L / 100$.
The averaging scales are $R = 1000 T_L, 100 T_L, 10 T_L, T_L, T_L/10$, which
correspond to the curves from top to bottom.} 
\end{figure}
We now check whether or not
the random variables (\ref{x2}) and (\ref{dx2}) follow
lognormal distributions by doing numerical simulation of the OU process
(\ref{ou}).
We use the numerical method called exact updating formula 
of the OU process proposed in Ref.\cite{g} with the parameters 
$x_0 = 0.0, T_L = 1.0, \kappa = 0.50, \Delta t = 1.0\times 10^{-3}$.
The integrals in Eqs. (\ref{x2}) and (\ref{dx2}) are numerically 
calculated as 
\begin{eqnarray}
 X^2_R(t_n) &=&
  \frac{1}{N_R}
  \sum_{j = -N_R / 2}^{N_R / 2}
   X(t_{n + j})^2, \\
\delta X^2_{r, R}(t_n) &=&
 \frac{1}{N_R}
  \sum_{j = -N_R / 2}^{N_R / 2}
  [X(t_{n + j} + t_{N_r})     - X(t_{n + j})]^2,\nonumber \\
\end{eqnarray}
where $N_R = R / \Delta t, N_r = r / \Delta t$ and $t_k = k \Delta t$.
Figures \ref{pdf} and \ref{pdfd} show probability density functions (PDF) of 
$X^2_R$ and $\delta X^2_{r, R}$ with or without taking logarithm
for various averaging scales $R$. The number of samples for the $R=1000 T_L$
case is $1.34 \times 10^6$ (for smaller $R$ cases, the number
is much larger).
For the largest values of $R = 1000 T_L$ shown here,
the PDFs are close to Gaussian
distributions as a consequence of the central limit theorem.
As we decrease the averaging scale $R$ to the correlation scale $T_L$,
the distributions of $X^2_R$ and $\delta X^2_{r, R}$ deviate 
from the Gaussian distribution. In contrast,
the log variables $\ln X^2_R$ and $\ln \delta X^2_{r, R}$
remain to be nearly Gaussian as shown in Figs.\ref{pdf}(b) and \ref{pdfd}(b).
Hence $X^2_R$ and $\delta X^2_{r, R}$ are lognormally distributed 
in this range of $R$.
This behavior is similar to the turbulence data analyzed in Ref.\cite{mht}. 
In addition, for the squared increments $\delta X^2_{r, R}$ with various $r < T_L$,
qualitatively same results are obtained.

\begin{figure}
\centerline{
\includegraphics[scale=0.38]{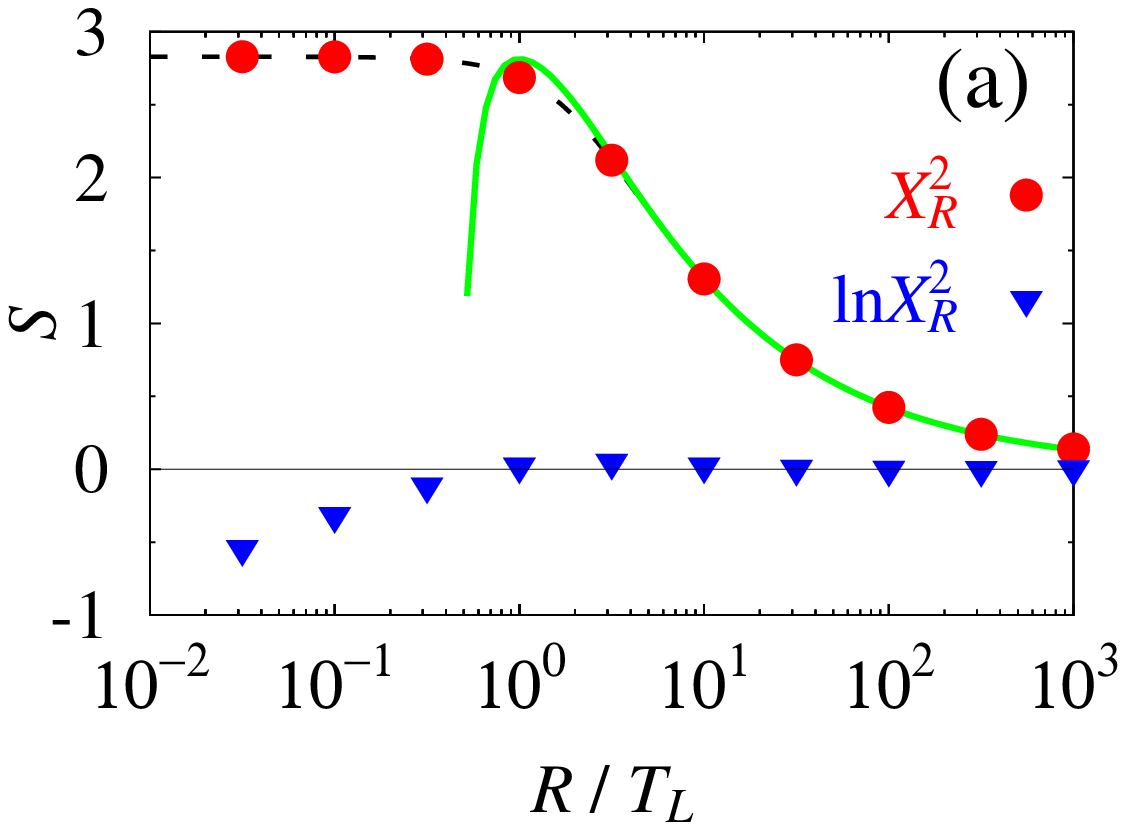}
\includegraphics[scale=0.38]{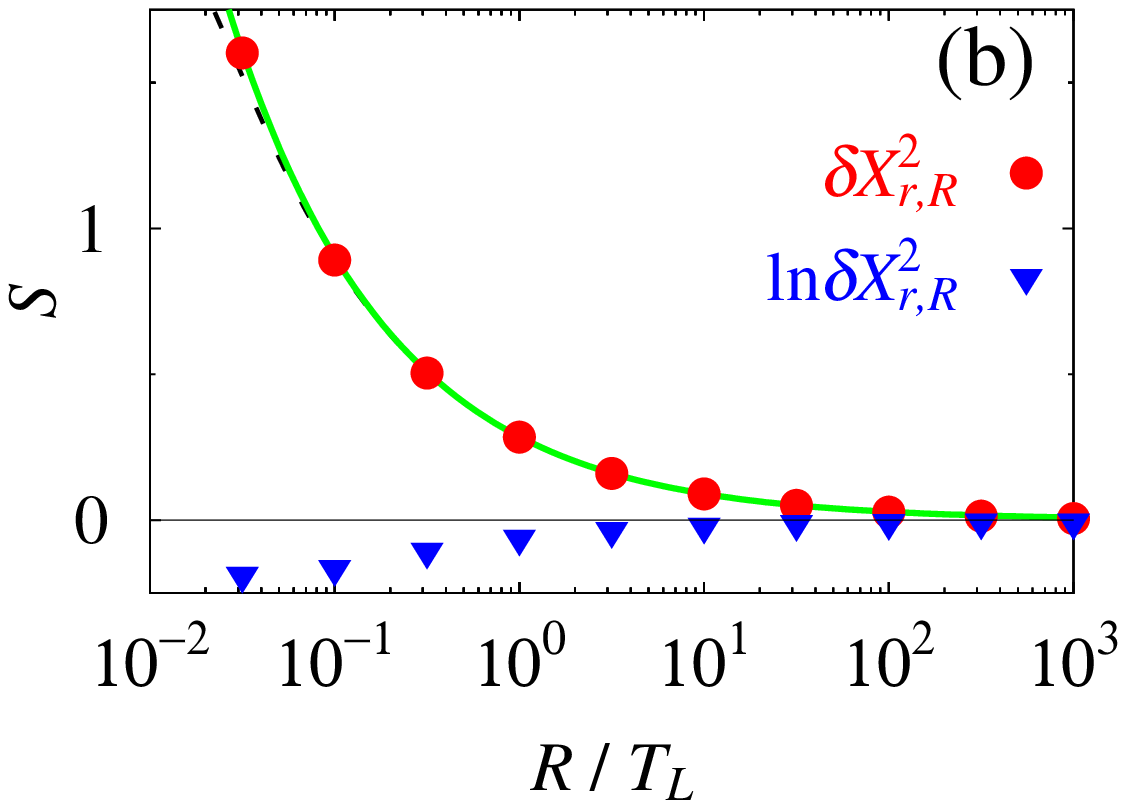}  
}
\vspace*{0.2cm} 
\centerline{ 
\includegraphics[scale=0.38]{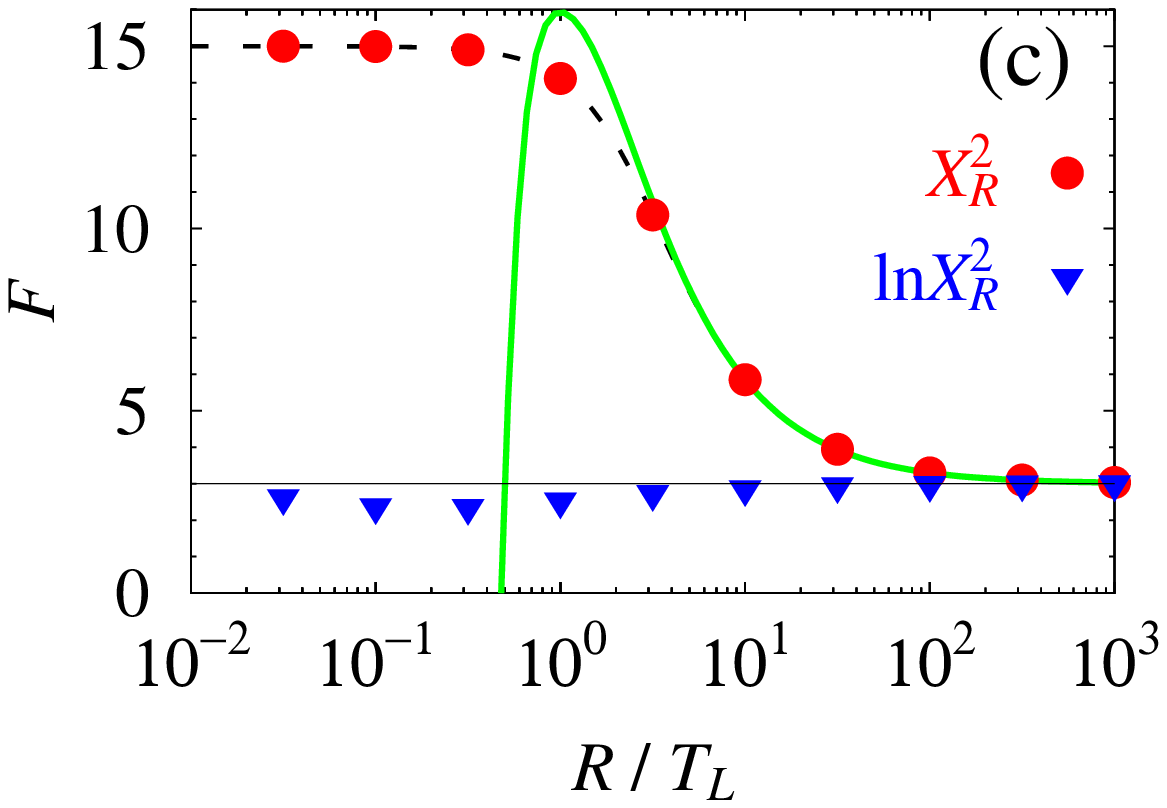}
\includegraphics[scale=0.38]{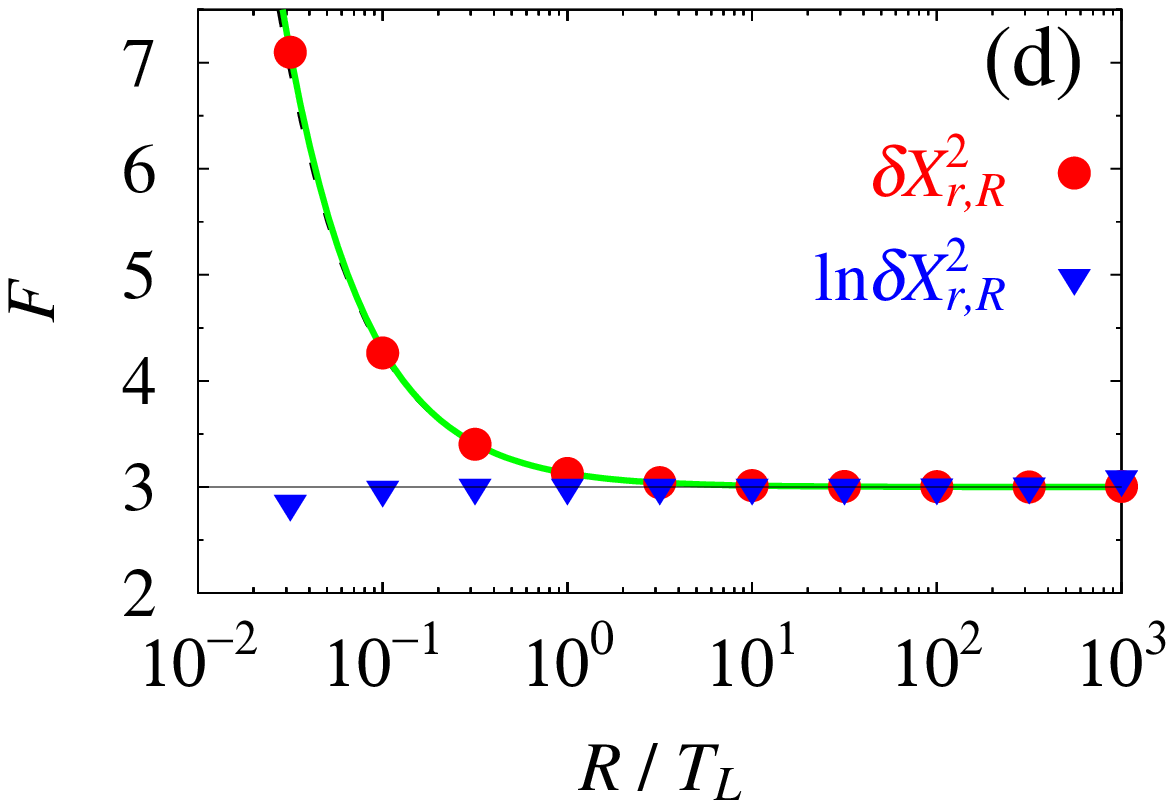}  
} 
\caption{\label{moments} (Color online)
The skewness of $X^2_R$ (a) and of $\delta X^2_{r, R}$ ($r = T_L/100$) (b); 
the flatness of $X^2_R$ (c) and of $\delta X^2_{r, R}$ (d) with or without logarithm.
They are plotted versus the averaging scale $R$.
The horizontal lines correspond to the skewness and the flatness
of the Gaussian distributions.
The thick curves are
asymptotic expressions of the moments of $X^2_R$ and $\delta X^2_{r, R}$,
Eqs.(\ref{s2}--\ref{f2}) and (\ref{sd2}--\ref{fd2}). The dashed curves
correspond the exact moment expressions (not involving any asymptotic argument).
}
\end{figure}

The approach to the lognormal distribution can be observed more quantitatively 
by looking at how the skewnesses and flatnesses of the log variables $\ln X^2_R, \ln \delta X^2_{r, R}$ 
change as functions of $R$ (the skewness $S(z)$ and flatness $F(z)$ of a random variable $z$ are defined
as $S(z) = \langle (z - \langle z \rangle)^3 \rangle / [V(z)]^{3/2}, 
F(z) = \langle (z - \langle z \rangle)^4 \rangle / [V(z)]^{2}$, where $V(z)$ 
is the variance $\langle (z - \langle z \rangle)^2 \rangle$).
In Fig.~\ref{moments}, it is seen that the moments of the variables with logarithm 
$\ln X^2_{R}, \ln \delta X^2_{r, R}$
approach already around $R = T_L$ to the values of the Gaussian distributions $(S,\, F) = (0,\, 3)$
whereas the moments of the variables without logarithm are still different from them. 
This is consistent with the behavior observed in Figs.\ref{pdf} and \ref{pdfd}.

\begin{figure}
\centerline{
\includegraphics[scale=0.38]{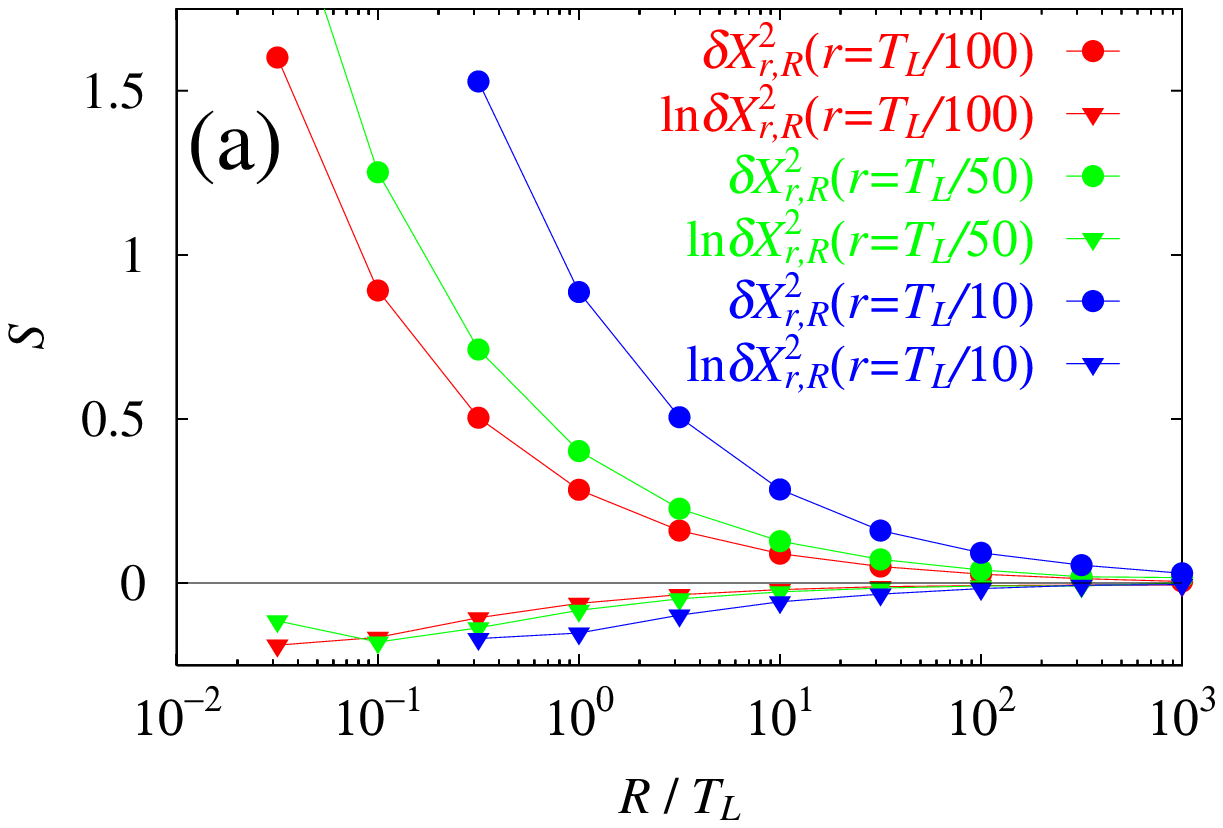}
\includegraphics[scale=0.38]{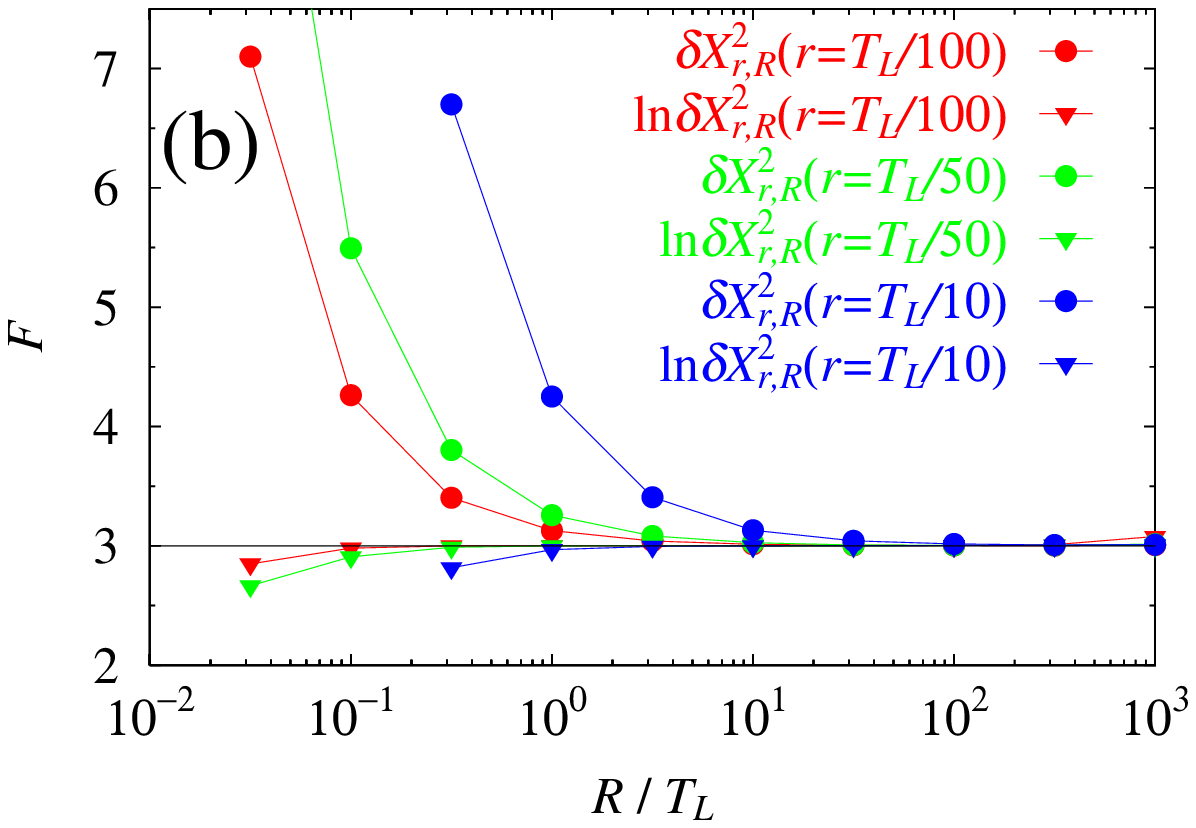}  
}
\caption{\label{varir} (Color online)
The skewness (a) and flatness (b) of the increments 
$\delta X^2_{r, R}$ and $\ln \delta X^2_{r, R}$
for three different values of $r$: $r = T_L/10,\, T_L/50,\, T_L/100$.
}
\end{figure}

For the increments $\delta X^2_{r, R}$, the skewness and the flatness
can depend on the difference $r$. Indeed a clear $r$-dependence is seen
in Fig.\ref{varir}.
However, the fast convergence to the Gaussian
of the $\ln \delta X^2_{r, R}$ around $R \sim L$ is not
affected by this dependence.
In fact, these graphs of the different $r$s can be collapsed to one curve
by normalizing $R$ with a different correlation scale from $T_L$ as we will show at
the end of the next section.

In summary, we observe that the large-scale lognormality holds also 
for the OU process as in the turbulence case \cite{mht}.
Further details on the behavior of the skewness and flatness of the OU process
are studied analytically in the next section.

\section{Analytical expressions of the moments of the Ornstein-Uhlenbeck process}
\label{s:analy}
The OU process is amenable to analytical calculation in many ways. 
However it is difficult to determine analytically
the PDF's of the quantities of our interest, Eqs.(\ref{x2}) and (\ref{dx2}),
or all their moments, since they involve integrations.
Instead we calculate the low-order moments, namely the skewness and 
the flatness of $\XRb$ and $\dXRb$. For them, full analytic results
are obtained but we present only the asymptotic expressions of them 
for $R / L \gg 1$ since they are sufficient for our purpose.

Then we compare them with the expressions
of the skewness and flatness of the lognormal distribution with 
the mean $e$ and the variance $v$:
\begin{eqnarray}
 S_{\LN} &=& \sqrt{\rho} (3 + \rho), \label{Sln} \\
 F_{\LN} &=& 3 + \rho (16 + 15 \rho + 6\rho^2 + \rho^3),\label{Fln}
\end{eqnarray}
where $\rho = v / e^2$.
As we have seen, the PDF's of the variables $\XRb, \dXRb$ become close to
the lognormal distributions for $R/T_L \sim 1$.
We study this behavior
by comparing the skewness and flatness of $\XRb, \dXRb$ with $S_{\LN}$ and $F_{\LN}$.
However, as we shall see below, it turns out that small corrections 
exist, which may be hard to be detected in numerical calculations 
in the previous section.

In our analytical calculation of the moments,
we first put the formal solution of the
Langevin equation (\ref{ou})
\begin{eqnarray}
 X(t) = \e^{-t/T_L}x_0 + \int_0^{t} \e^{-(t - s)/T_L} \sqrt{2\kappa} \xi(s) d s.
\label{fs}    
\end{eqnarray}
into the coarse-grained variables Eqs.(\ref{x2}) and (\ref{dx2}). 
For the correlation or the variance,
we take a suitable powers of them and take the ensemble average with respect
to the Langevin noise by using Eq.(\ref{xi}). 
So that we can reduce their calculations into multiple integrals of certain exponential functions.
For the skeweness and the flatness, we use more compact integral representations
proposed in Ref.\cite{rice}, which is explained in Appendix.
To calculate the resultant integrals
we use a symbolic computation software, Maple. 
The integrated result still contains dependence on time $t$. For example, the mean
of $\XRb$ reads
\begin{eqnarray}
 \XR = \kappa T_L
       \left[
         1 +
	 \frac{T_L}{2R}
	 (\e^{-R/T_L} - \e^{R/T_L})~ \e^{-2t/T_L}
       \right].
\end{eqnarray}
Remember that this time $t$ is much larger than 
the correlation time $T_L$ to be consistent to the simulations in Sec.\ref{s:num}.
We hence regard $\e^{-t/T_L} = 0$.
From now on, we drop the dependency on $t$ of
all the moments of $\XRb$ and $\dXRb$ calculated here.
This does not affect our study since the OU process
becomes steady for $t \gg T_L$.

We now list the expressions of the moments. 
Here we write $\Lambda = R / T_L$ for brevity.
For $\XRbwt$, the mean, variance, skewness, and flatness are respectively,
\begin{eqnarray}
 E(\XRbwt) &=& \kappa T_L, \\
 V(\XRbwt) &=& (\kappa T_L)^2 \left[ 2\Lambda^{-1} + \Lambda^{-2} (\e^{-2\Lambda} - 1) \right]
 ,\\
 S(\XRbwt) &=& 12 W_0^{-3/2} [\e^{-2\Lambda} - 1 + \Lambda (\e^{-2\Lambda} + 1)] 
 \label{Xs},\\
 F(\XRbwt) &=& 3 + 6 W^{-2}_0
              [ \e^{-4\Lambda} + 28\e^{-2\Lambda} -29 \nonumber \\
      &&                   + 16\Lambda^2\e^{-2\Lambda} + 40\Lambda \e^{-2\Lambda} + 20\Lambda
			   ],\label{Xf}
\end{eqnarray}
where $W_0 = 2\Lambda^{-1} + \Lambda^{-2} (\e^{-2\Lambda} - 1)$.
For the coarse grained increment $\dXRbwt$, by writing $r / T_L = \lambda$, 
we give the results in the leading
order of $\Lambda^{-1} = (R / T_L)^{-1}$ to avoid lengthy expressions:
\begin{eqnarray}
E(\dXRbwt) &=&  2\kappa T_L \e^{-\lambda}(\e^{\lambda} - 1),\\
V(\dXRbwt) &=& (2\kappa T_L)^2 \Lambda^{-1} \e^{-2\lambda}
             \big[
	       3 \e^{2\lambda}
	       - 4(\lambda + 1) \e^{\lambda} \nonumber \\
	   &&    + 2\lambda + 1
	     \big], \\ 
S(\dXRbwt) &=& 3\Lambda^{-2} \e^{-3\lambda} W_1^{-3/2}
                \big[10 \e^{3\lambda} 
		- 5(\lambda^2 + 3\lambda + 3)\e^{2\lambda} \nonumber \\
            &&    + 2(4\lambda^2 + 6\lambda + 3) \e^{\lambda}
	     -(3\lambda^2 + 3\lambda + 1)\big] ,\label{ds}\\
F(\dXRbwt) &=& 3+ 
             \Lambda^{-3} \e^{-4\lambda} W_1^{-2} 
           [
	     525 \e^{4\lambda} \nonumber \\
	 &&    - (56 \lambda^3 + 336 \lambda^2 + 840 \lambda + 840) \e^{3\lambda} \nonumber \\
         &&    + (224 \lambda^3 + 672 \lambda^2 + 840 \lambda + 420) \e^{2\lambda} \nonumber \\
         &&    - (216 \lambda^3 + 432 \lambda^2 + 360\lambda + 120) \e^{\lambda} \nonumber \\
         &&    + (64\lambda^3 + 96\lambda^2 + 60\lambda + 15)
	    ],\label{df}
\end{eqnarray}
where $W_1 = \Lambda^{-1} \e^{-2\lambda} [3 \e^{2\lambda} - 4(\lambda + 1) \e^{\lambda} + 2\lambda + 1]$.
Notice that we obtain analytic expressions of $S$ and $F$, which
are shown as the dashed curves in Fig.\ref{moments}.

With these asymptotic expressions,
we next re-write the skewness $S$ and flatness $F$
as functions of the variance over the squared mean $\rho = V/E^2$
to compare them with the lognormal ones $S_{\LN}$ and $F_{\LN}$.

For $X_R^2$, in the limits of $\Lambda = R / T_L \to \infty$,
we have
\begin{eqnarray}
  S(\XRbwt) &\simeq& 0 + \sqrt{\rho}\left( 3 \quad\quad - \frac{3}{16}\rho^2   \right) \label{s2},\\
  F(\XRbwt) &\simeq& 3 + \rho \left(15 + \frac{3}{8}\rho - \frac{39}{16}\rho^2 \right)\label{f2},
\end{eqnarray}   
where
\begin{eqnarray}
  \rho &=& \frac{V(\XRbwt)}{E(\XRbwt)^2} \simeq \frac{2T_L}{R}-\left(\frac{T_L}{R}\right)^2.
   \label{rho}
\end{eqnarray}
The asymptotic expression Eqs.(\ref{s2}) and (\ref{f2})
agree with data shown in Fig.\ref{moments} for $R \ge \sqrt{10} T_L$.

For the velocity increments $\delta X_{r, R}^2$,
we have $\rho = V / E^2 = \Lambda^{-1}\big[ 3 \e^{2\lambda} - 4(\lambda + 1)
\e^{\lambda} \nonumber + 2\lambda + 1 \big]/(\e^{\lambda}-1)^2$.
By taking the limit of $\lambda = r / T_L \to 0$, we obtain
expressions
\begin{eqnarray}
  S(\dXRbwt) &\simeq& 0 + \sqrt{\rho} \times \frac{99}{40}, \label{sd2}\\
  F(\dXRbwt) &\simeq& 3 + \rho \times \frac{1359}{140}, \label{fd2}
\end{eqnarray}
where
\begin{eqnarray}
  \rho =\frac{V(\dXRbwt)}{E(\dXRbwt)^2} \simeq \frac{4 \lambda}{3 \Lambda} = \frac{4 r}{3 R}.
   \label{rhod}   
\end{eqnarray}
The equations (\ref{sd2}) and (\ref{fd2}) agree with the data shown in Fig.\ref{moments}
for $R \ge 10^{-3/2} T_L$ (covering all data points), thanks to the small value
$r = 10^{-2}T_L$.

We now compare the analytic expressions of the OU quantities
Eqs.(\ref{s2}--\ref{f2}) and (\ref{sd2}--\ref{fd2}) with 
the log-normal ones Eqs.(\ref{Sln}--\ref{Fln}).
We can observe the following:
(i) Indeed as $R \to \infty ~(\rho \to 0)$, $S$ and $F$ of
both $\XRbwt$ and $\dXRbwt$
tend to the values of Gaussian distribution, $S = 0$ and $F = 3$.
Before reaching this state, we see an approach to the lognormal 
distribution.
(ii) The sub-leading terms of $S$ and $F$ have the same powers of 
$\rho$ as $S_{\LN}$ and $F_{\LN}$,
which is in favor of the large-scale lognormality of the OU process.
However, the constant in the subleading term is slightly different
from those of the lognormal distribution.
In this sense, the coarse-grained quantities $\XRbwt$ and $\dXRbwt$ of the
OU process becomes nearly lognormally distributed when $R / T_L$ is 
large but not exactly so.

\begin{figure}
\centerline{
\includegraphics[scale=0.38]{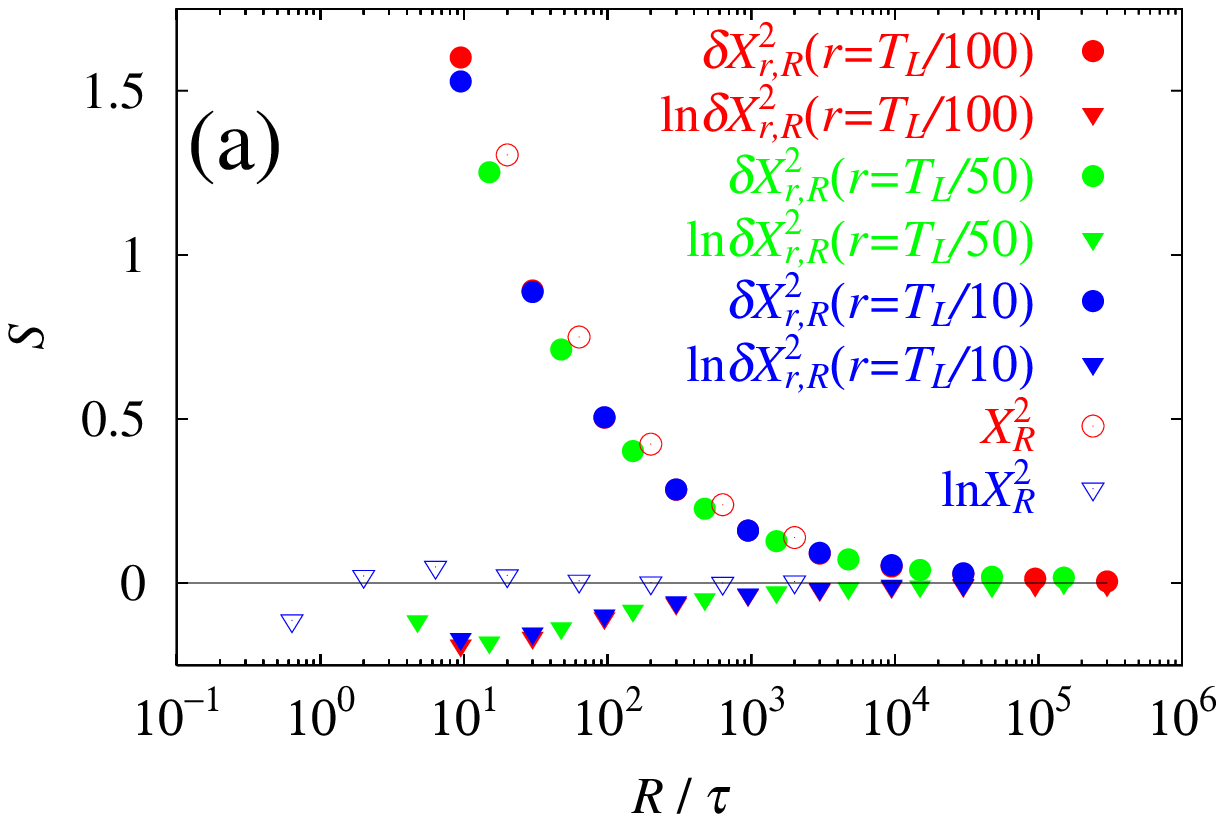}
\includegraphics[scale=0.38]{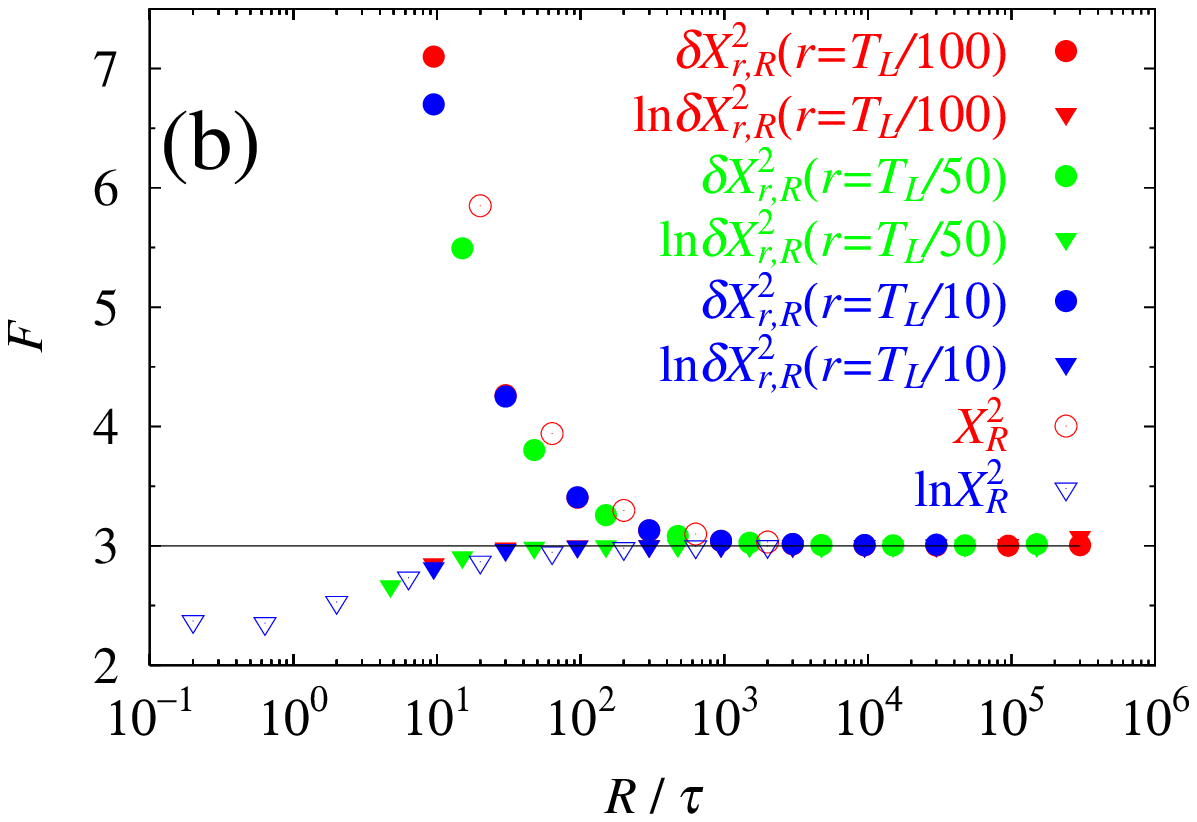}  
}
\caption{\label{collapse} (Color online)
The collapsed skewness (a) and flatness (b) of $\XRbwt$ and $\dXRbwt$
as functions of $R$ normalized with $\tau(X^2)$ and $\tau(\delta X^2_r)$
defined in (\ref{tau}) and (\ref{taud}).
}
\end{figure}

So far we focus on how the moments vary as a function of $R / T_L$, where
$R$ is normalized by the correlation scale $T_L$ of the OU process.
We close this section with a digression by pointing out another normalization
scale of the problem. This is indicated by the above analytic results.
For $\XRbwt$, let us go back to the definition Eq.(\ref{x2}). The integral
scale of the integrand of Eq.(\ref{x2}) is
\begin{eqnarray}
\tau(X^2) &=&
 \frac{\int_0^{\infty} \langle [X^2(t) - \langle X^2 \rangle][X^2(t + s)- \langle X^2 \rangle] \rangle d s}
      {\langle [X^2(t) - \langle X^2 \rangle]^2\rangle} \nonumber \\
 &=& \frac{T_L}{2}. \label{tau}
\end{eqnarray}
For $\dXRbwt$,  the integral
scale of its integrand defined similarly is calculated as
\begin{eqnarray}
\tau(\delta X^2_r)
 = T_L \frac{3\e^{2\lambda} - 4 ( \lambda + 1 ) \e^\lambda + 2\lambda + 1}{4(\e^{\lambda} - 1)^2}
 \simeq
 \frac{r}{3} \label{taud}
\end{eqnarray}
for small $\lambda = r / T_L$.
Notice that the $\rho$ variables defined in Eqs.(\ref{rho}) and (\ref{rhod})
contains the corresponding integral scales of the integrands (\ref{tau}) and (\ref{taud}),
namely, $\rho = 4\tau / R$ for both $\XRbwt$ and $\dXRbwt$.
Hence it implies that a suitable normalization scale of $R$
for collapsing the moment data plotted
in Figs.\ref{moments} and \ref{varir} are respectively 
the integral scales $\tau(X^2)$ and $\tau(\delta X^2_r)$ of the integrands.
Indeed this normalization
yields a better collapse as shown in Fig.\ref{collapse} for both $\XRbwt$ 
and $\dXRbwt$ with different $r$.
This suggests that, when we look into a coarse-grained quantity over scale $R$
of a function $f_R(y) = (1/R)\int_{y}^{y + R} f(y') d y'$ of a correlated fluctuation $y$, 
the correlation scale (integral scale) of $f(y)$ in certain cases can be a better characteristic scale 
than the correlation scale of the fluctuation $y$
itself. 
We believe that this is the case not only for the OU process but also for
a general fluctuation with a correlation.
However this normalization blurs the fact that the large-scale lognormality
of the OU process occurs at $R/T_L \sim 1$.
This kind of normalization will be reported elsewhere.

\section{Summary and discussion}
\label{s:dcr}
This study was motivated by the large-scale lognormality of turbulence
that was recently observed experimentally in grid, boundary-layer, 
and jet turbulences \cite{mht}.
In this lognormality, the correlation scale plays a pivotal role. Namely
when the averaging scale to the correlation scale are of order unity,
the averaged squared velocity and velocity increments become 
lognormally distributed fluctuations. We here anticipated that
this large-scale lognormality is a property of correlated random variables
in general.

To test this idea, we took the Ornstein-Uhlenbeck process as
a simplest means to generate correlated random variables.
Our numerical simulation indicates that the OU process
also exhibits the large-scale lognormality for about the same range of
$R / L$ as observed in turbulence. Then, to further investigate the approach to the lognormal 
distribution, we calculated analytically the skewness and
flatness of the coarse-grained quantities of the OU process.
The first subleading term of the asymptotic expressions for large $R / L$ of the moments
revealed
that indeed the moments behave nearly as those of the log-normal distribution.
However there is small deviation from the lognormal distribution.

Now we speculate on behavior of the higher order moments than forth order.
The analytical calculation of them for the coarse-grained OU variables becomes
so complicated that we did not try even with Maple.
For the lognormal variable $x_{\rm LN}$, the moments can be written
with $\rho$, the ratio of the variance to the squared mean, as
\begin{eqnarray}
 H^{(q)}_{\rm LN}
  &=& \frac{\langle [x_{\rm LN} - \langle x_{\rm LN} \rangle]^q \rangle}{[V(x_{\rm LN})]^{q/2}}   \nonumber \\
  &=& \rho^{-\frac{q}{2}} \sum_{k=0}^{q}\binom{q}{k}(\rho + 1)^{\frac{1}{2}k(k - 1)}(-1)^{q - k}, \label{binom}
\end{eqnarray}
where $\binom{q}{k}$ is the binomial coefficients.
By expanding Eq.(\ref{binom}) with $\rho$, the moment can be written as
\begin{eqnarray}
 H^{(q)}_{\rm LN}
  &=& \begin{cases}
        0 + s_1^{(q)} \rho^{\frac{1}{2}} + s_2^{(q)} \rho^{\frac{3}{2}} + \ldots + \rho^{\frac{1}{2}q(q-2)} ~(\mbox{$q$: odd}), \cr
        (q - 1)!! + f_1^{(q)} \rho + f_2^{(q)} \rho^{2} + \ldots + \rho^{\frac{1}{2}q(q-2)}  ~(\mbox{$q$: even}). \cr       
      \end{cases}
      \label{lnh}
\end{eqnarray}
Here the terms of the negative powers of $\rho$ are shown to vanish
and $s_1^{(q)}, f_1^{(q)}, \ldots$ are suitable coefficients, which can be written
with the binomial coefficients. 
The $\rho^0$ terms in Eq.(\ref{lnh}), namely $0$  and $(q - 1)!! = (q-1)(q-3) \cdots 5 \cdot 3 \cdot 1$,
correspond to the values of the Gaussian distribution.
We speculate that, for the coarse-grained quantities of the OU variable,
their higher-order moments have the similar expression as Eq.(\ref{lnh}).
As we saw in the previous section, the behavior for small $\rho$ (corresponding to $R / T_L$ being larger than unity)
is relevant for the large-scale lognormality.
Their first term should coincide with that of Eq.(\ref{lnh}) because of the usual central limit theorem. 
It is tempting to speculate also that the second term has the same power of $\rho$ but the
value of the the coefficient, $s_1^{(q)}$ or $f_1^{(q)}$ are different.
We have no idea how much they are different. 

In relation to the real turbulence data, the OU process is not a good
representation as a whole since, for example, it does not have deviation from being Gaussian
(intermittency effect) and equivalence of the energy cascade nor the energy
dissipation rate. However, as long as we focus on the large-scale fluctuations
of turbulence where the single point velocity or the velocity increments are
close to being Gaussian, the OU process serves a useful and analytically
tractable model. In this study we regarded that the correlation at the integral
scale is the most important aspect to be modeled by the OU process.

Here we have seen that the correlation plays the essential role
in the near lognormal behavior of the coarse-grained positive 
quantities Eqs.(\ref{x2}) and (\ref{dx2}). 
Roughly speaking, this near lognormality around the correlation scale 
may be regarded as an intermediate state, or ``a rule of thumb''
before  
the central limit theorem holds with much larger averaging scale $R$.
For the turbulence cases studied in \cite{mht}, similar mechanism is 
likely to work.
Other examples of lognormal behavior involving a coarse-graining average
include cosmological density fluctuations (see e.g., \cite{sc}), which
may have a similar structure as studied here.

\section{Appendix}
Here we explain briefly how we calculate the moments of the coarse-grained quantities.
In particular we use the Rice's method given in Section 3.9 of Ref.\cite{rice} to reduce the number of the multiple
integrals. For illustration of the method, let us take as an example the second-order moment of
$X_R^2(t)$
\begin{eqnarray}
\langle [X^2_R(t)]^2 \rangle
 = \frac{1}{R^2}
 \int_{t -R/2}^{t + \frac{R}{2}} d t_1
 \int_{t -R/2}^{t + \frac{R}{2}} d t_2
 ~\langle X^2(t_1)  X^2(t_2) \rangle.\nonumber \\
\end{eqnarray}
The idea of the Rice's method is to write the moment
$\langle X^2(t_1) X^2(t_2) \rangle$ in terms
of the correlation function $\langle X(t_1) X(t_2) \rangle = \psi(t_2 - t_1)$
by noticing that $X(t_1)$ and $X(t_2)$ are multivariate Gaussian variables
whose covariace matrix is known completely.
In principle, any moment of $X(t_1), X(t_2), \ldots, X(t_n)$ can be expressed
with $\psi(t_i - t_j)$.
Specifically, with the correlation function for large $t$
\begin{equation}
\psi(\tau)= \langle X(t) X(t + \tau) \rangle = \kappa T_L \e^{-|\tau|/T_L},
\end{equation}
the variance and other central moments can be written as follows:
\begin{eqnarray}
 \langle [X^2_R(t)- \langle X^2_R\rangle]^2 \rangle 
&=& \frac{2}{R^2}\int_{t-\frac{R}{2}}^{t+\frac{R}{2}} d t_1 
                \int_{t-\frac{R}{2}}^{t+\frac{R}{2}} d t_2 \psi^2(t_1-t_2) \nonumber \\
&=&\frac{4}{R^2}\int_0^R(R-x)\psi^2(x) d x \nonumber \\
&=&\frac{\kappa^2T_L^2}{4R^2}\left[2T_LR+T_L^2(\e^{-2R/T_L}-1)\right] \nonumber \\
\end{eqnarray}
\begin{eqnarray}
 \langle [X^2_R(t)- \langle X^2_R \rangle]^3 \rangle 
&=&\frac{8}{R^3}\int_{t-\frac{R}{2}}^{t+\frac{R}{2}}d t_1 
                \int_{t-\frac{R}{2}}^{t+\frac{R}{2}}d t_2 
                \int_{t-\frac{R}{2}}^{t+\frac{R}{2}}d t_3 \nonumber \\
          &&    \times  \psi(t_1-t_2)\psi(t_2-t_3)\psi(t_3-t_1)
\nonumber\\
&=& \frac{48}{R^3}\int_0^R dx (R-x)\psi(R-x) \nonumber \\
 && \times 
    \int_0^x d y \psi(x-y)\psi(y)
 \label{slast}
\end{eqnarray}
\begin{eqnarray}
 \langle [X^2_R(t) - \langle X^2_R\rangle]^4\rangle
&=& 
3\langle (X^2_R(t)- \langle X^2_R \rangle)^2 \rangle^2  \nonumber \\
&& \hspace*{-2cm} +\frac{48}{R^4}\int_{t-\frac{R}{2}}^{t+\frac{R}{2}}d t_1 
                 \int_{t-\frac{R}{2}}^{t+\frac{R}{2}}d t_2 
                 \int_{t-\frac{R}{2}}^{t+\frac{R}{2}}d t_3
                 \int_{t-\frac{R}{2}}^{t+\frac{R}{2}}d t_4 \nonumber \\
&& \hspace*{-2cm} \times   \psi(t_2-t_1)\psi(t_3-t_1)\psi(t_4-t_2)\psi(t_4-t_3)
\nonumber\\
&& \hspace*{-2.5cm} = 3\langle (X^2_R(t)- \langle X^2_R \rangle )^2 \rangle^2
\nonumber\\
&& \hspace*{-2cm} +\frac{768}{R^4}
\int_0^R d x(R-x)\int_0^R d y\psi(y)\psi(x-y) \nonumber \\
&& \times \int_0^y d z\psi(x-z)\psi(z).
\label{flast}
\end{eqnarray}
Here we change variables in the integrals by using the symmetry of $\psi$
to reduce the double integral to a single integral (for details, see \cite{rice}). 
The integrals
(\ref{slast}) and (\ref{flast}) are calculated analytically with the software
Maple. 

Concerning the increments, we write its correlation function for large $t$
\begin{eqnarray}
\psi_r(\tau) &=& \langle \delta_rX(t+\tau)\delta_rX(t) \rangle \nonumber \\
 &=&  \kappa T_L \left[ \e^{-|\tau|/T_L}(2-\e^{-r/T_L})-\e^{-|r-|\tau||/T_L} \right].
\nonumber \\
\end{eqnarray}
Using the same argument as in the case of $X_R(t)$, 
we can express the central moments with the correlation $\psi_r$ as follows:
\begin{eqnarray}
 \langle [\delta X^2_{r,R}(t)- \langle \delta X^2_{r,R}(t) \rangle]^2 \rangle
&=&\frac{4}{(R-r)^2}\int_0^{R-r}(R-r-x)
 \nonumber \\
&& \times \psi_r^2(x)d x
\nonumber\\
 \langle [\delta X^2_{r,R}(t)- \langle \delta X^2_{r,R}(t) \rangle]^3 \rangle
&=&\frac{48}{(R-r)^3}\int_0^{R-r} d x(R-r-x) \nonumber \\
&& \times \int_0^x d y \psi_r(x-y)\psi_r(x)\psi_r(y) \nonumber \\
\label{dslast}
\end{eqnarray}
\begin{eqnarray}
 \langle [\delta X^2_{r,R}(t)\!&-&\! \langle \delta X^2_{r,R}(t) \rangle]^4 \rangle
\nonumber\\
&=&3  \langle [\delta X^2_{r,R}(t)- \langle \delta X^2_{r,R}(t) \rangle ]^2 \rangle^2
\nonumber\\
&& +\frac{768}{(R-r)^4}
 \int_0^{R-r} d x(R-r-x) \nonumber \\
&& \hspace*{-1cm} \times  \int_0^{R-r} d y \psi_r(y)\psi_r(x-y) 
 \int_0^y d z\psi_r(x-z)\psi_r(z).\nonumber \\
\label{dflast}
\end{eqnarray}
The final forms (\ref{dslast}) and (\ref{dflast}) are calculated with Maple.

\section*{Acknowledgments}
We gratefully acknowledge discussion with H.~Mouri.
We are supported by the Grant-in-aid for Scientific Research C (22540402) from
the Japan Society for the Promotion of Sciences.
T.M. is supported by the Grant-in-Aid for the Global COE Program
"The Next Generation of Physics, Spun from Universality and Emergence"
from the MEXT of Japan.

\end{document}